\documentclass[11pt]{article}

\usepackage{arxiv}

\usepackage[utf8]{inputenc} 
\usepackage[T1]{fontenc}
\usepackage{amsmath,amsfonts,amssymb,bm}
\usepackage{graphicx,color}
\usepackage{setspace}
\doublespace   
\usepackage{fancyhdr}
\usepackage[absolute,overlay]{textpos}
\usepackage{multicol}
\usepackage[authoryear,round]{natbib}
\usepackage{booktabs} 
\usepackage{caption}
\usepackage{subcaption}
\usepackage{url}
\usepackage{hyperref}
\usepackage{float}
\usepackage{algorithm2e}
\usepackage{adjustbox}

\definecolor{Gray}{gray}{0.9}

\allowdisplaybreaks

\usepackage{arydshln}
\makeatletter
\renewcommand*\env@matrix[1][*\c@MaxMatrixCols c]{%
	\hskip -\arraycolsep
	\let\@ifnextchar\new@ifnextchar
	\array{#1}}
\makeatother

\RequirePackage{doi}

\usepackage{bbm} 

\newcommand{\tCollapsedurl}{\if0\blind
	{
		\url{\tCollapsedurl}
	} \fi
	\if1\blind
	{ 
		\url{github.com/}\texttt{[url redacted in blinded version]}
	} \fi}

\title{Copula-based models for correlated circular data}

\author{
Francesco Lagona\\
    \scriptsize{Dpt. of Political Sciences}\\
    \scriptsize{Roma Tre University}\\
    \scriptsize{\texttt{francesco.lagona@uniroma3.it}}
\And
    Marco Mingione\\
    \scriptsize{Dpt. of Political Sciences}\\
    \scriptsize{Roma Tre University}\\
    \scriptsize{\texttt{marco.mingione@uniroma3.it}}
}

\begin{document}
\maketitle
\begin{abstract}
We exploit Gaussian copulas to specify a class of multivariate circular distributions and obtain parametric models for the analysis of correlated circular data. This approach provides a straightforward extension of traditional multivariate normal models to the circular setting, without imposing restrictions on the marginal data distribution nor requiring overwhelming routines for parameter estimation. The proposal is illustrated on two case studies of animal orientation and sea currents, where we propose an autoregressive model for circular time series and a geostatistical model for circular spatial series.   
\end{abstract}

\section{Introduction}
Correlated circular data arise when multiple angular measurements are taken on the same sample unit and are encountered across several research areas. Examples include ecological studies of animal movement \citep{hodel2022, ranalli2020model}, biometrical studies of torsional angles in molecules \citep{greco2023finite, Mardia_etal:2008}, environmental studies of sea currents \citep{wang2014modeling} and social studies of crime intensity \citep{gelfand2017:crime}. 
These data are often referred to as hypertoroidal data, as they can be represented as points in the $K$-dimensional torus (hypertorus for short), which is the Cartesian product of multiple copies of the unit circle. 

The statistical analysis of correlated circular data requires the specification of multivariate distributions with hypertoroidal support. Existing proposals of hypertoroidal distributions include the multivariate von Mises distribution \citep{Mardia_etal:2008}, the multivariate wrapped normal distribution \citep{Nodehi_etal:2021}, the multiple trigonometric sums distribution \citep{FernándezDurán_etal:2014} and the multivariate projected normal distribution \citep{Mastrantonio:2018,Wang2013}.

Here, we introduce a class of hypertoroidal distributions that rely on Gaussian copulas. Compared to the aforementioned proposals, this class does not impose restrictions on the marginal distribution of the data nor requires overwhelming estimation routines for likelihood-based inference. 

The idea of using copulas in directional statistics has already been explored since the seminal work by \cite{Wehrly_etal1980}, where a class of bivariate distribution on the torus was defined by binding two circular marginal densities using a circular copula, nowadays known as a {\em circula}. Circulas however lack both flexibility and scalability. On the one hand, when a circula integrates two wrapped Cauchy marginals, an elegant toroidal distribution is obtained \citep{Kato_Pewsey2015}, but when alternative marginals are exploited, circulas do not necessarily result in particularly attractive bivariate
circular models \citep{Jones_etal2015}. On the other hand, extending circulas from a bivariate to a multivariate setting is not straightforward. 

Gaussian copulas do not suffer from these restrictions
and are especially attractive for modelling correlated circular data because the copula structure is not much affected by the data dimension. Moreover, the dependence structure is specified in terms of a correlation matrix, allowing the extension of traditional multivariate normal models to the circular setting, and simultaneously satisfying the periodic requirements of hypertoroidal distributions.     

\section{Copula-based toroidal distributions}
\label{sec:bivariate}
We conveniently illustrate our proposal under the bivariate setting, postponing the straightforward multivariate extension to Section \ref{sec:multivariate}. Accordingly, let $\bm{u}=(u_1, u_2)$ be a sample drawn from a bivariate uniform distribution with support $(0,1)^2$. A mapping $C: (0,1)^{2}\rightarrow (0,1)$ is called a copula if it is a cumulative distribution function (cdf) with uniform marginals. Copulas are often specified by relying on known distributions. For example, let $\Phi$ and $\Phi_{\bm{\Omega}}$ indicate the cdf of the standard normal $N(0,1)$ and the bivariate normal $N(\bm{0}, \bm{\Omega})$, respectively, where $\bm{\Omega}$ is the correlation matrix, known up to a correlation parameter $\rho$, namely
\[\bm{\Omega}=\begin{pmatrix}
    1 & \rho \\
    \rho & 1
\end{pmatrix}.\] 
The Gaussian copula is a bivariate cdf with uniform marginals, defined as 
\begin{equation}\label{eq:copula}
C_{\bm \Omega}(\bm{u})=\Phi_{\bm{\Omega}}(\Phi^{-1}(u_1), \Phi^{-1}(u_2)),\end{equation}
where $\Phi^{-1}$ is the quantile function of the standard normal distribution. 

Gaussian copulas can be conveniently exploited to specify the joint cdf of a pair of two absolutely continuous random variables $(Y_1,Y_2)$, with marginal cdf $F_1(y_1)$ and $F_2(y_2)$. Specifically, the Sklar's theorem \citep{sklar1959} guarantees that the function  
\begin{align}\label{eq:toroidal.cdfA}
    F(y_1,y_2)=C_{\bm \Omega}(F_1(y_1), F_2(y_2)) = \Phi_{\bm{\Omega}}\left(\Phi^{-1}( F_1(y_1)),  
\Phi^{-1}(F_2(y_2))\right)
\end{align}
is the cdf of a bivariate random variable $(Y_1,Y_2)$, with marginals $F_1(y_1)$ and $F_2(y_2)$. 

Let $\mathbb{C}$ be the unit circle. To introduce the case of interest here, we assume that $Y_1$ and $Y_2$ are two circular random variables with cdf $F_1(y_1)$ and $F_2(y_2)$, $y_1,y_2 \in \mathbb{C}$. The cdf of a circular random variable $Y$ can be evaluated from an arbitrary starting
point on $\mathbb{C}$, for example from $\mu-\pi$, where $\mu$ is the circular mean of $Y$. By the integral transformation theorem, the function
\begin{equation}
\label{eq:transformation}
z_k(y) =\Phi^{-1}(F_k(y)), \qquad y \in \mathbb{C}, \quad k=1,2\end{equation} 
 transforms a circular variable with cdf $F_k$ into a standard normal, while its inverse
\[z_k^{-1}(z)=F_k^{-1}(\Phi(z)) \qquad z \in \mathbb{R}, \quad k=1,2\]
transforms a standard normal into a circular variable with cdf $F_k$.
By setting $\bm{z}(\bm{y})=(z_1(y_1),z_2(y_2))$, the bivariate cdf in \eqref{eq:toroidal.cdfA} reduces to $F(y_1,y_2)=\Phi_{\bm{\Omega}}\left(\bm{z}(\bm{y})\right)$.

\begin{figure}
\centering
\begin{subfigure}{.45\textwidth}
    \centering
    \includegraphics[width=.9\textwidth]{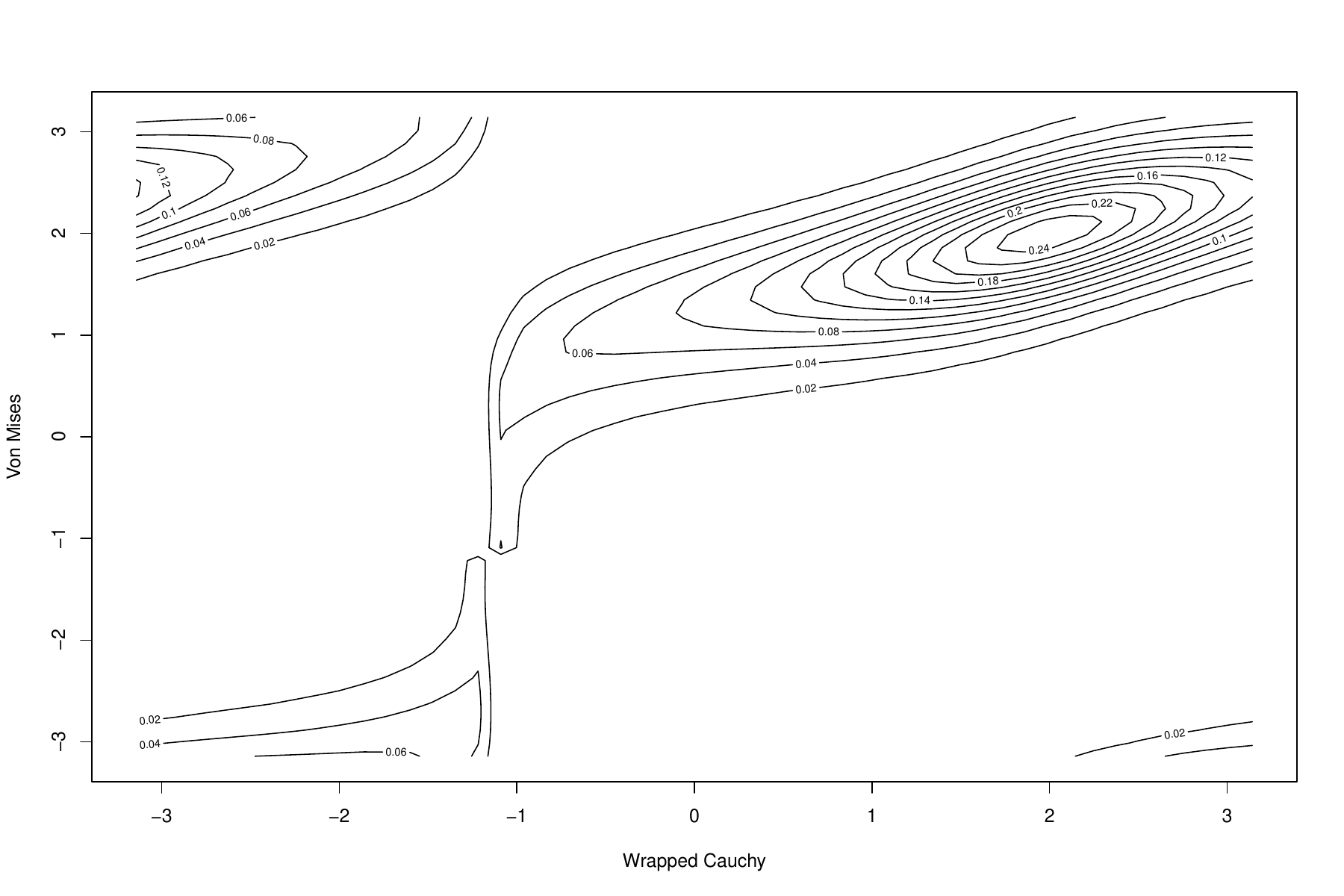}
    \caption{}
\end{subfigure}
\begin{subfigure}{.45\textwidth}
    \centering
    \includegraphics[width=.9\textwidth]{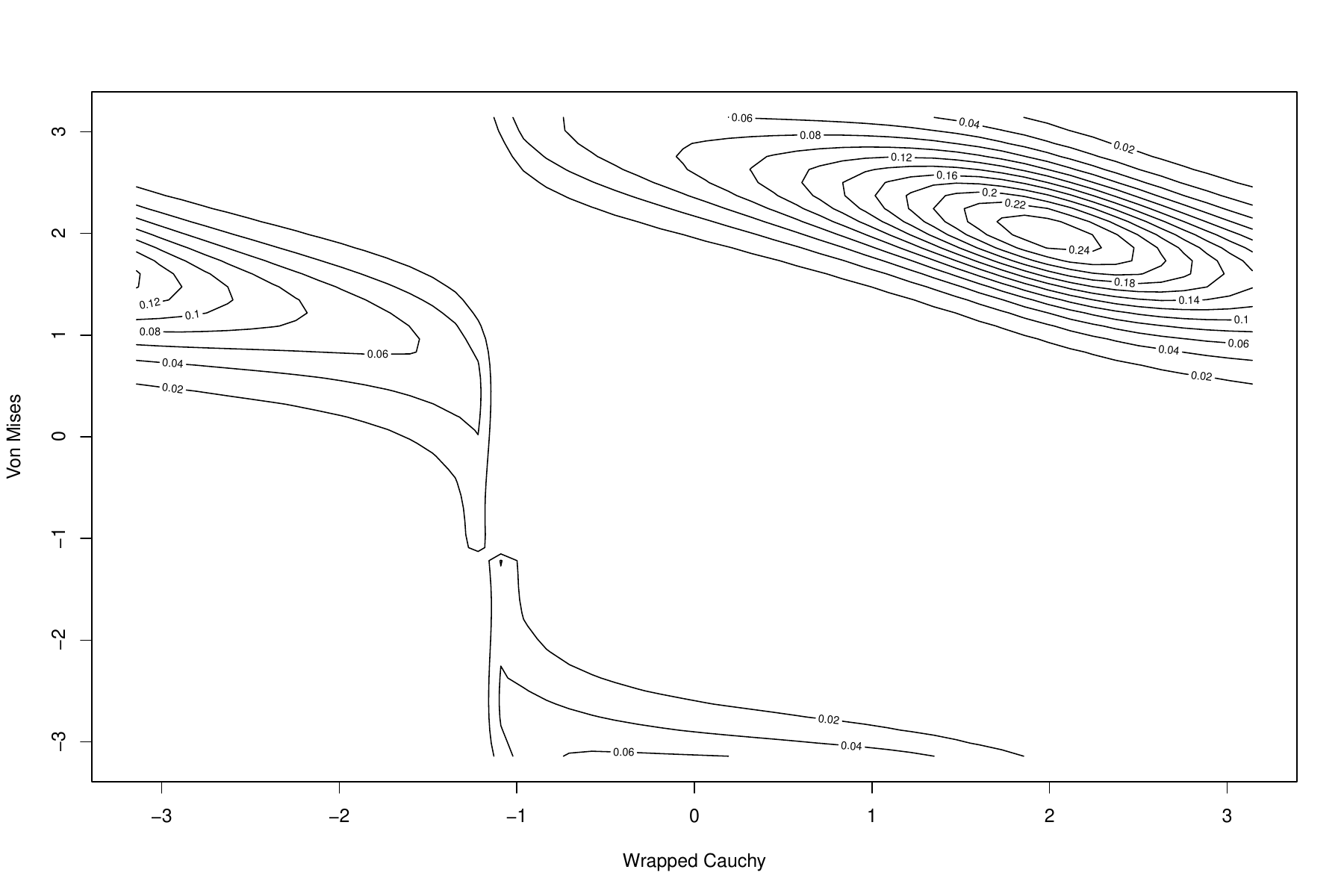}
    \caption{}
\end{subfigure}
\caption{A copula-based toroidal distribution obtained by binding a  wrapped Cauchy with parameters $\mu=2$ and $\kappa=0.3$ (x-axis) and a von Mises with parameters $\mu=2$ and $\kappa=2$ (y-axis), and correlation parameter $\rho=0.8$ (a) and $\rho=-0.8$ (b).}
\label{fig:examplecontour}
\end{figure}

Differentiation of $F(y_1,y_2)$ yields a toroidal density with support $\mathbb{C}^2$. To illustrate, let $\phi$ and $\phi_{\bm{\Omega}}$ be the densities of the standard normal $N(0,1)$ and $N(\bm{0}, \bm{\Omega})$, respectively. Also, let $f_1(y_1)$ and $f_2(y_2)$ be the marginal densities of $Y_1$ and $Y_2$. Under this setting, $\bm{z}(\bm{Y})\sim N(\bm{0}, \bm{\Omega})$ and, by the chain rule, the joint density of $(Y_1, Y_2)$ is given by     
\begin{align} 
\label{eq:toroidal.pdf}
&f(y_1,y_2)= \frac{\partial^2}{\partial y_1 \partial y_2}\Phi_{\bm{\Omega}}(\bm{z}(\bm{y})) \nonumber\\
=&\phi_{\bm{\Omega}}(\bm{z}(\bm{y}))\frac{1}{\phi(z_1(y_1))\phi(z_2(y_2))}f_{1}(y_1)f_2(y_2) \nonumber\\
=&\mid \bm{\Omega}\mid ^{-1/2}\exp \left(\frac{1}{2}\bm{z}(\bm{y})^{\sf T}\left(\bm{I}-\bm{\Omega}^{-1}\right)\bm{z}(\bm{y})\right)f_{1}(y_1)f_2(y_2) \nonumber\\
=&c_{\bm{\Omega}}(\bm{z}(\bm{y}))f_{1}(y_1)f_2(y_2),
\end{align}

where the copula density $c_{\bm{\Omega}}(\bm{z}(\bm{y}))=c_{\bm{\Omega}}(z_1(y_1),z_2(y_2))$ fulfills the periodicity condition, i.e.
\[c_{\bm{\Omega}}(z_1(y_1),z_2(y_2))= c(z_1(y_1\pm 2l \pi),z_2(y_2\pm 2m \pi)),\]
with $l, m \in \mathbb{N}, \forall (y_1,y_2) \in \mathbb{C}^2$. In particular, when $\bm{\Omega}$ is equal to the identity matrix, then  \eqref{eq:toroidal.pdf} reduces to the joint product density of two independent circular variables. Otherwise, $\bm{\Omega}$ accommodates
both positive and negative dependence through the correlation parameter $\rho$. Figure \ref{fig:examplecontour} displays two toroidal densities, obtained by binding a wrapped Cauchy distribution
$$
f_1(y_1)=\frac{1}{2\pi}\frac{1-\kappa^2}{1+\kappa^2-2\kappa \cos(y_1-\mu)}, \quad y_1 \in \mathbb{C}
$$
and a von Mises distribution
$$
f_2(y_2)=\frac{\exp(\kappa \cos(y_2 - \mu))}{2\pi I_0(\kappa)}, \quad y_2 \in \mathbb{C}
$$
with $\rho=\pm 0.8$. 

While Sklar's theorem guarantees that the two marginal densities of \eqref{eq:toroidal.pdf} are $f_1(y_1)$ and $f_2(y_2)$, the straightforward proof of the following theorem shows that the conditional distribution of $Y_2$ given $y_1$, say $f(y_2 \mid y_1)$, is obtained by perturbing the marginal circular density $f_2(y_2)$ by a weighting function that identically reduces to 1 under independence.

\paragraph{Theorem 1.}\label{theorem1}
Let $$f(y_1,y_2)=c_{\bm{\Omega}}(z_1(y_1),z_2(y_2))f_{1}(y_1)f_2(y_2)$$
be the joint density of a toroidal distribution with marginals $f_k(y_k), \, y_k \in \mathbb{C}, \, k = 1, 2$ and correlation matrix 
$$\bm{\Omega}=\begin{pmatrix}
    1 & \rho \\
    \rho & 1
\end{pmatrix}.$$  
The conditional density of $Y_2$ given $y_1$ is given by    
\begin{equation}\label{eq:cond.density}f(y_2 \mid y_1)=f_2(y_2)\frac{\phi_{\bm{\Omega}}(z_2(y_2) \mid z_1(y_1))}{\phi(z_2(y_2))},\end{equation}
where $\phi_{\bm{\Omega}}(z_2(y_2) \mid z_1(y_1))$ is a normal distribution with mean $\rho z_1(y_1)$ and variance $1-\rho^2$.

\paragraph{Proof of Theorem 1}
By relying on \eqref{eq:toroidal.pdf}, the conditional density of interest is obtained as
\begin{align*}f(y_2 \mid y_1)=&\frac{f(y_1,y_2)}{f_1(y_1)}\\
=& \frac{\phi_{\bm{\Omega}}(z_1(y_1), z_2(y_2))}{\phi(z_1(y_1))\phi(z_2(y_2))}f_2(y_2)\\
=& f_2(y_2)\frac{\phi_{\bm{\Omega}}(z_2(y_2) \mid z_1(y_1))}{\phi(z_2(y_2))}.
\end{align*}
Recalling that $\bm{z}(\bm{Y})\sim N(\bm{0}, \bm{\Omega})$, the conditional density $\phi_{\bm{\Omega}}(z_2(y_2) \, \mid \,z_1(y_1))$ is normal with mean $\rho z_{1}(y_1)$ and variance $1-\rho^2$.

Furthermore, the $p$-quantile of \eqref{eq:cond.density}, say $q_{Y_2 \mid y_1}(p)$, can be conveniently obtained by a simple transformation of normal quantiles. Specifically, let $q_p$ be the $p$-quantile of the conditional normal density $\phi_{\bm{\Omega}}(z_2(y_2) \mid z_1(y_1))$. Under this setting,   
\begin{align*}p= &
P\left(z_2(Y_2)<q_p \big| y_1\right)\\
=&P\left(z_2^{-1}\left(z_2(Y_2)\right)<z_2^{-1} \left(q_p\right) \big| y_1\right)\\
=&P\left(Y_2< z_2^{-1}(q_p) \big| y_1\right),
\end{align*}
or alternatively,
$q_{Y_2 \mid y_1}(p)= z_2^{-1}(q_p)$. By computing this conditional quantile at $p=0.025$ and $p=0.975$, we obtain a circular 95\% prediction arc for $Y_2$ given $y_1$.  

\section{Copula-based hypertoroidal distributions}
\label{sec:multivariate}
Here, we illustrate the multivariate extension of the proposal described in Section \ref{sec:bivariate}.
Let $\bm{\Omega}$ be a $K\times K$ correlation matrix and let $f_1(y_1), \ldots, f_K(y_K)$ be $K$ circular densities. Then, a hypertoroidal density with marginals $f_k(y_k), k=1, \ldots, K$ and support $\mathbb{C}^K$ can be specified as   
\begin{equation}
\label{eq:multivariate.pdf}
f(\bm{y})=\mid \bm{\Omega}\mid ^{-1/2}\exp \left(\frac{1}{2}\bm{z}(\bm{y})^{\sf T}\left(\bm{I}-\bm{\Omega}^{-1}\right)\bm{z}(\bm{y})\right)\prod_{k=1}^{K}f_{k}(y_k),
\end{equation}
where $\bm{z}(\bm{y})=(z_1(y_1),\ldots, z_K(y_K))$, and 
\begin{equation}\label{eq:mult.transformation} 
z_k(y_k)=\Phi^{-1}(F_{k}(y_k)), \quad k=1,\ldots, K.
\end{equation}

Without loss of generality, the multivariate extension of Theorem 1 can be obtained by partitioning the vector $\bm{z}(\bm{y})$, say \[\bm{z}(\bm{y})=(z_1(y_1),\bm{z}_2(\bm{y}_2)),\] where $\bm{z}_2(\bm{y}_2)=(z_{2}(y_2), \ldots z_{K}(y_K))$ and considering the associated partitioning of $\bm{\Omega}$, namely
\[\bm{\Omega}=\left(\begin{array}{cc}
     1 &  \bm{\omega}_{12} \\
     \bm{\omega}_{21} & \bm{\Omega}_{22}
\end{array}\right),\]
 where $\bm{\omega}_{12}=\bm{\omega}_{21}^{\sf T}$ is the vector of the correlations between $z_1(y_1)$, and $\bm{z}_2(\bm{y}_2)$ and $\bm{\Omega}_{22}$ is the marginal correlation matrix of $\bm{z}_2(\bm{Y}_2)$. Under this setting,  
\[f(y_1 \mid y_2, \ldots, y_K)=\frac{f_1(y_1)}{\phi(z_1(y_1))}\phi_{\bm{\Omega}}(z_1(y_1) \mid \bm{z}_2(\bm{y}_2)),\]
where $\phi_{\bm{\Omega}}(z_1(y_1) \mid \bm{z}_2(\bm{y}_2))$ is a normal distribution with mean $\bm{\omega}_{21}^{T}\bm{\Omega}_{22}^{-1}\bm{z}_2(\bm{y}_2)$ and variance $1-\bm{\omega}_{21}^{T}\bm{\Omega}_{22}^{-1}\bm{\omega}_{21}$. Accordingly, the conditional $p$-quantile of $Y_1$ given the rest of the sample is given by
\begin{equation}\label{eq:mult.cond.quantile}
q_{Y_1 \mid y_2 \ldots, y_K}(p)=z_1^{-1} (q_p),
\end{equation}
where $q_p$ is the quantile of the conditional normal density $\phi_{\bm{\Omega}}(z_1(y_1) \mid z_2(y_2), \ldots z_K(y_K))$. By computing this conditional quantile at $p=0.025$ and $p=0.975$, we obtain a circular 95\% prediction arc for $Y_1$ given the rest of the sample.

\section{Parametric inference}
\label{sec:estimation}
Parametric models can be specified by assuming that the correlation matrix $\bm{\Omega}$ and the marginal densities $f_k(y_k)$ in \eqref{eq:multivariate.pdf} are known up to a set of parameters to be estimated, say $\bm{\Omega}=\bm{\Omega}(\bm{\rho})$ and $f_k(y_k)=f_k(y_k; \bm{\theta}_k)$, leading to a parametric version of \eqref{eq:mult.transformation}, i.e.
$z_k(y_k)=z_k(y_k; \bm{\theta}_k)=\Phi^{-1}(F_{k}(y_k; \bm{\theta}_k))$.
These yield a parameter-indexed version of \eqref{eq:multivariate.pdf}, i.e.
\begin{align}
\label{eq:general.model}
f(\bm{y};  \bm{\theta}, \bm{\rho})=&\mid \bm{\Omega}(\bm{\rho})\mid ^{-1/2}\nonumber\\ &\cdot \exp \left(\frac{1}{2}\bm{z}(\bm{y}; \bm{\theta})^{\sf T}\left(\bm{I}-\bm{\Omega}^{-1}(\bm{\rho})\right)\bm{z}(\bm{y}; \bm{\theta})\right)\nonumber \\ & \cdot\prod_{k=1}^{K}f_{k}(y_k; \bm{\theta}_k),
\end{align}
where $\bm{\theta} = (\bm{\theta}_1, \ldots, \bm{\theta}_K)$. If $\bm{y}=(\bm{y}_1 , \ldots, \bm{y}_n )$ indicates a sample of multivariate observations independently drawn from \eqref{eq:general.model}, maximum likelihood estimation reduces to maximization of the log-likelihood
\begin{equation}
\label{eq:loglik}
l(\bm{\theta}, \bm{\rho})=\sum_{i=1}^{n}\log f(\bm{y}_i; \bm{\theta}, \bm{\rho})=a( \bm{\theta}, \bm{\rho})+b(\bm{\theta}),
\end{equation}
where
\begin{align*}
a(\bm{\rho}, \bm{\theta})=&
-\frac{n}{2}\mid \bm{\Omega}(\bm{\rho})\mid \nonumber \\ & + \sum_{i=1}^{n}\left(\frac{1}{2}\bm{z}(\bm{y}_i; \bm{\theta})^{\sf T}\left(\bm{I}-\bm{\Omega}^{-1}(\bm{\rho})\right)\bm{z}(\bm{y}_i; \bm{\theta})\right),\\
b(\bm{\theta})=&\sum_{i=1}^{n}\sum_{k=1}^{K}\log f_{k}(y_{ik}; \bm{\theta}_k).
\end{align*}
Maximization of \eqref{eq:loglik} is facilitated by initializing a standard optimization routine at a starting point $(\bm{\theta}_0, \bm{\rho}_0)$, obtained using an inference-from-margins approach (IFM; \cite{kim2016multivariate}). Under the proposed setting, the IFM approach reduces to find an initial estimate $\bm{\theta}_0$ by maximizing $b(\bm{\theta})$ and finding afterwards an estimate $\bm{\rho}_0$ by maximizing  $a(\bm{\theta}_0, \bm{\rho})$ with respect to $\bm{\rho}$. These values are used to initialize the PORT algorithm available in the \texttt{nlminb R} statistical software, but other routines could in principle be used. All codes to reproduce the results in Section \ref{application} are available at  \\ \texttt{https://github.com/minmar94/CopulaCircular}.

\section{Applications}\label{application}
We illustrate our proposal on two case studies of animal orientation and sea currents where multivariate circular data are respectively correlated in time and space.

\subsection{An auto-regressive model for animal orientation}
\begin{figure}
\centering
\begin{subfigure}[b]{.45\textwidth}
    \centering
    \includegraphics[width=.8\textwidth]{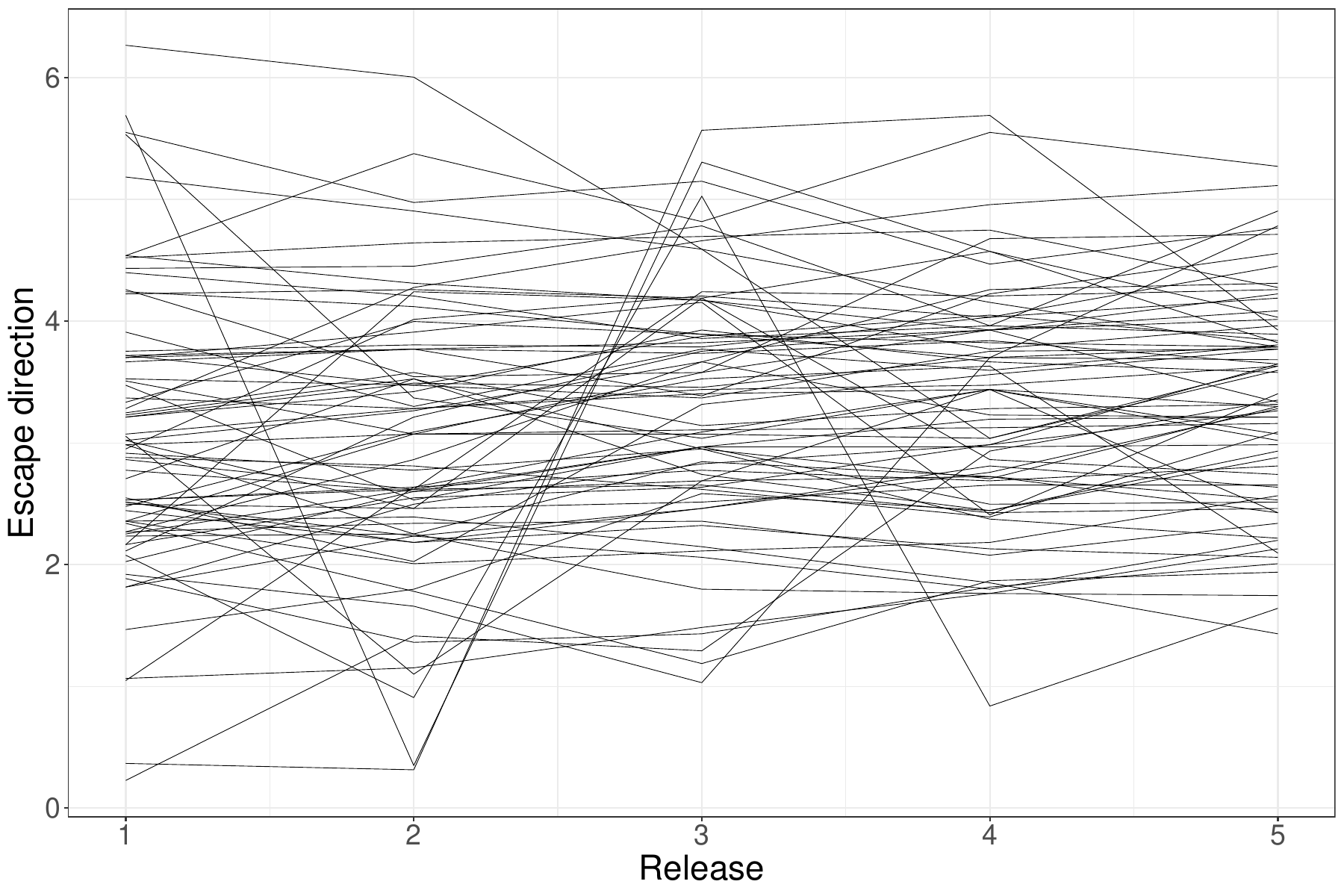}
    \caption{}
   \label{fig:sandhoopers.data}
\end{subfigure}
\begin{subfigure}[b]{.45\textwidth}
    \centering
    \includegraphics[width=.8\textwidth]{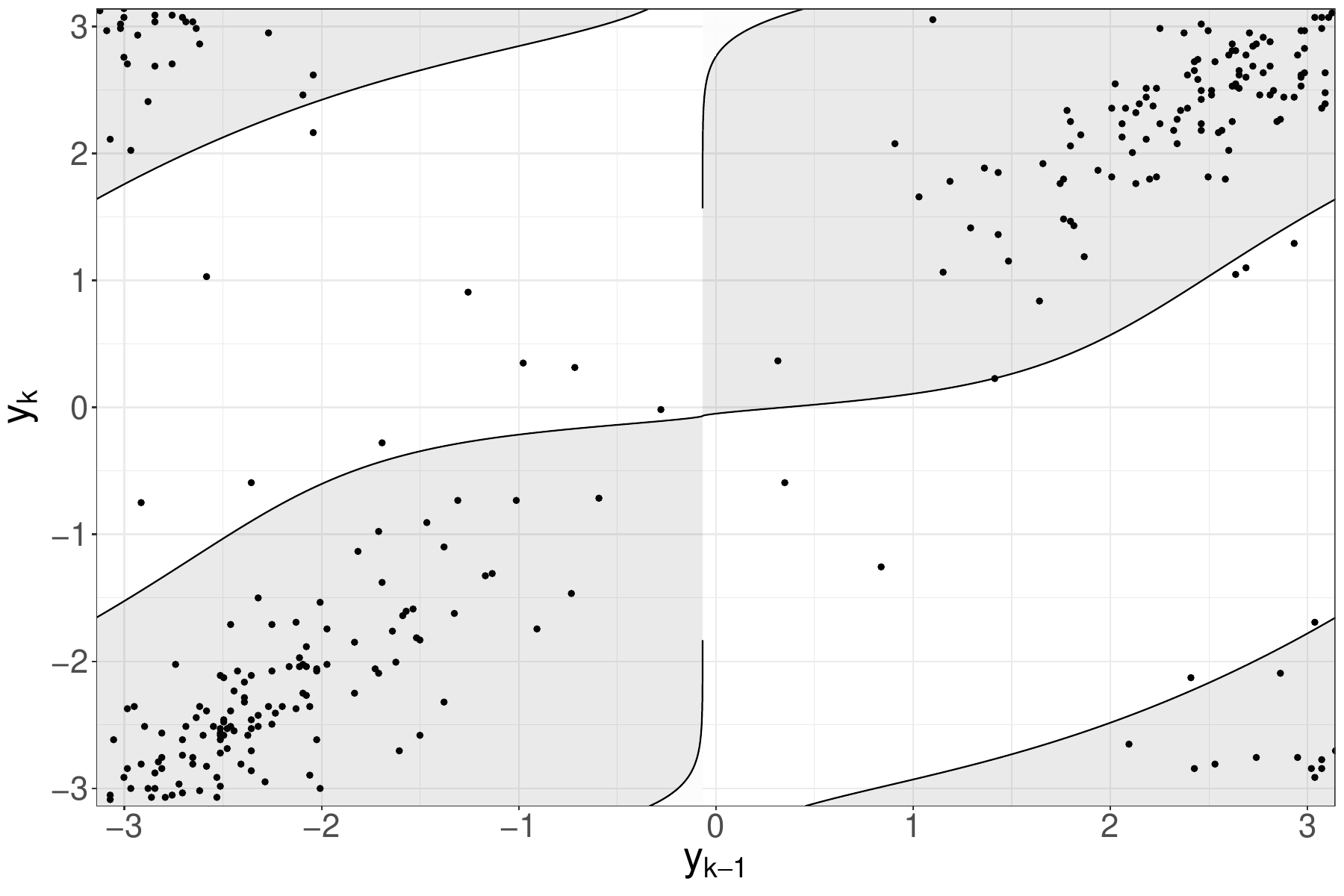}
    \caption{}
   \label{fig:esttorus}
\end{subfigure}
\begin{subfigure}[b]{.45\textwidth}
    \centering
\includegraphics[width=.8\textwidth]{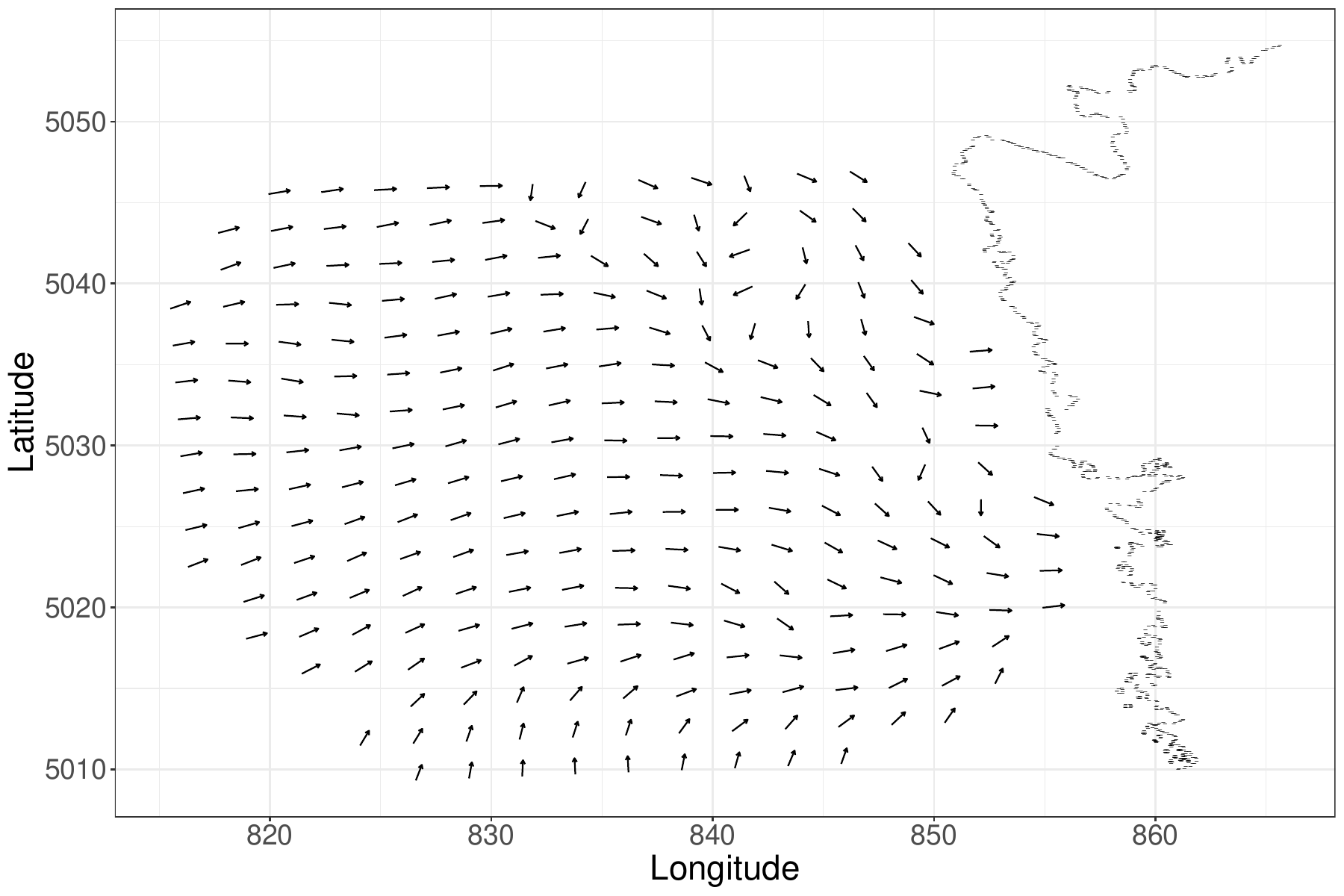}
\caption{}
\label{fig:sea.current.data} 
\end{subfigure}
\begin{subfigure}[b]{.45\textwidth}
    \centering
\includegraphics[width=.8\textwidth]{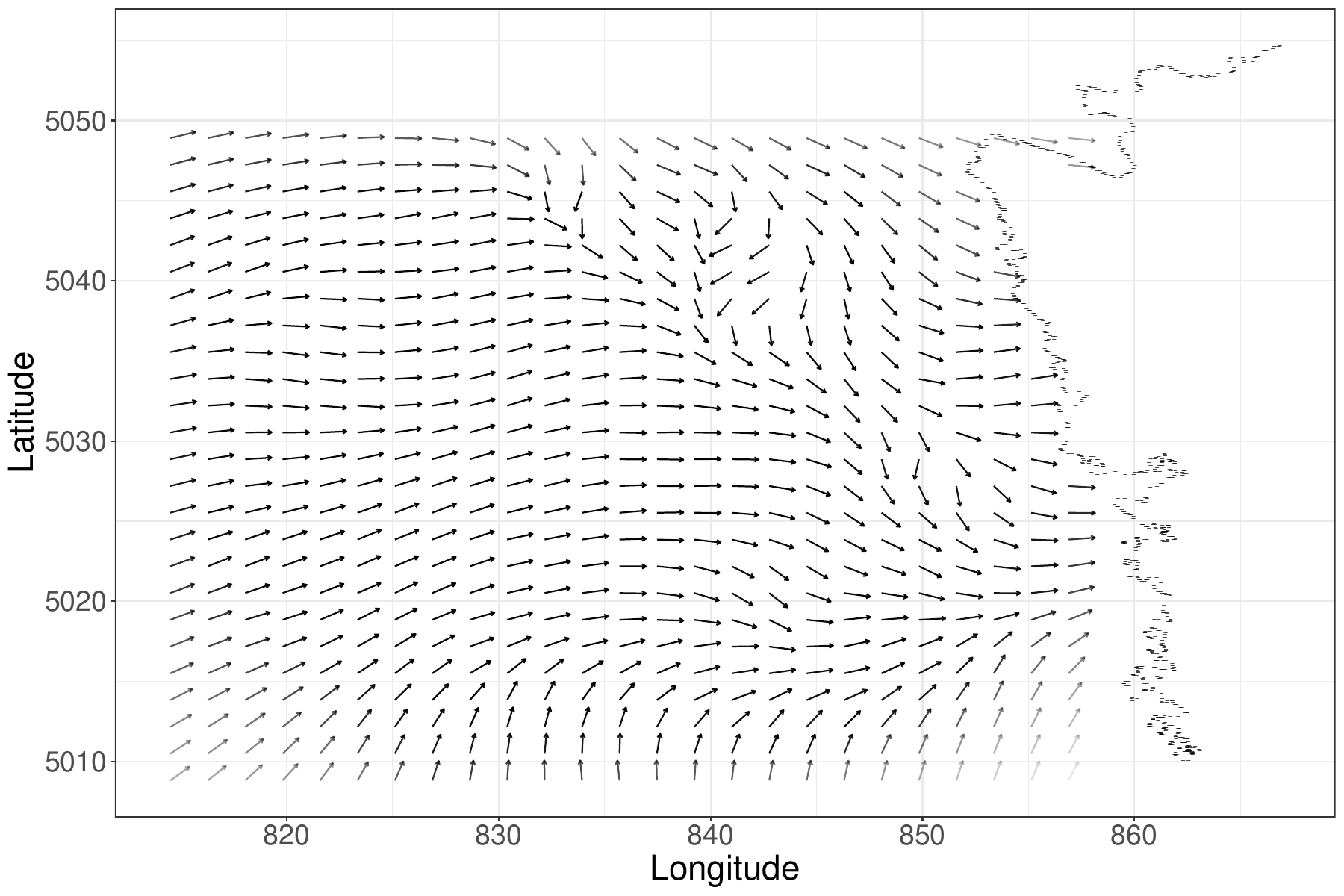}
\caption{}
\label{fig:kriging} 
\end{subfigure}
\caption{(a) Escape directions for sandhoppers over five consecutive releases; (b) 95\% prediction band of escape direction given the previous escape direction, as estimated by a circular autoregressive model; (c) sea current directions in the Northern Adriatic Sea, between the Gulf of Trieste and the Bay of Kvarner; (d) spatial prediction of sea current given the observed data, as obtained by a circular geostatistical model.}
\end{figure}

Figure \ref{fig:sandhoopers.data} displays the time series of the observed escape directions of 72 sandhoppers ({\it Talitrus saltator}), which were released sequentially on $K=5$ occasions. These data have been previously examined by circular mixed effect models \citep{NunezAntonio2014,Song2007book}, assuming that observations are conditionally independent given a subject-specific random intercept, or by circular auto-regressive models \citep{lagona2016regression} that rely on the multivariate von Mises distribution \citep{Mardia_etal:2008}.

Here, we investigate the dependence structure of the data by integrating a von Mises distribution known up to a circular mean parameter $\mu \in \mathbb{C}$ and a circular concentration parameter $\kappa>0$, with an AR(1) auto-regressive time series model, known up to a correlation parameter $\mid \rho \mid <1$, and specified by a correlation matrix $\bm{\Omega}$ with entries $\omega_{hk}=\rho^{\mid h-k \mid}$, $h,k=1, \ldots 5$.
Specifically, the data marginal distribution of the escape directions is captured by the von Mises density, while $\rho$ accounts for the temporal correlation of the data, transformed by $z(y; \mu, \kappa)=\Phi^{-1}(F(y; \mu, \kappa))$,
where $F(y; \mu, \kappa)$ is the cdf of a von Mises distribution with mean $\mu$ and concentration $\kappa$. Under this setting, \eqref{eq:general.model} reduces to a 3-parameter hypertoroidal distribution on $\mathbb{C}^5$, which assumes that escape directions are stationary and predictable by the previous observation. 

The first row of Table \ref{tab:estimates} displays the parameter estimates with related uncertainty, highlighting a significant correlation parameter and therefore motivating the choice of an auto-regressive model. Standard errors are obtained by inverting the Hessian matrix. Figure \ref{fig:esttorus} displays the 95\% confidence bands of the direction taken at the $k$-th release, given the value at the previous release, while dots indicate the observed pairs $(y_{i,k-1}, y_{ik})$ of the successive escape directions of each sandhopper $i = 1, \ldots, 72$. The band has been obtained by plugging the parameter estimates in \eqref{eq:mult.cond.quantile}, and setting $p \in \lbrace 0.025, 0.975\rbrace$. The periodic shape of the band can be interpreted by recalling that the grey area covers a portion of the surface of a torus.  

\begin{table*}[t]
    \centering
    \caption{Parameter estimates and standard errors for two copula-based toroidal models.}
    \label{tab:estimates}
    \begin{tabular}{lll|rrr}
         &&&Correlation & Mean & Concentration \\
         Case study & Domain & Model&  $\rho$ & $\mu$ & $\log(\kappa)$ \\
         \midrule
        Sandhoppers & Time &Autoregressive & 0.620 (0.036) & 3.074 (0.020) & 0.508 (0.081) \\
        Sea current & Space &Geostatistical & 0.559 (0.110) & 0.430 (0.019) & 1.313 (0.170)\\
        \bottomrule
    \end{tabular}
\end{table*}

\subsection{Marine currents}
\label{subsec:marine.currents}
Correlated circular data may arise not only in the form of time series but also in the form of spatial series. The second motivating example considered in this paper is a case study of spatial angular measurements. Figure \ref{fig:sea.current.data} displays the directions of surface sea currents obtained by high-frequency radars installed in the eastern part of the northern Adriatic. These data are part of the NASCUM (North Adriatic Surface Current Mapping) project \citep{mihanovic2011surface}. The figure includes observations taken on a sample of $K=245$ spatial locations, spreading up to 30km far from the coast,  between the Gulf of Trieste and the Bay of Kvarner. The coastlines
on the right side of the picture are the Croatian coasts of the Istrian peninsula.
These data have been previously examined by autoregressive spatial models that rely on the multivariate von Mises distribution \citep{Lagona_etalSERRA2015}. The multivariate von Mises distribution is however known up to an intractable normalizing constant and, as a result, parameter estimation is obtained by pseudo-likelihood \citep{Mardia_etal:2008} or MCMC-likelihood methods \citep{lagona2016regression}.  
Alternatively, the proposed copula-based approach offers a setting where maximum likelihood estimation is viable. To illustrate, let $\bm{y} = (y_1, \dots, y_K)$ be the vector of observed sea current values at location $1, \dots, K$. A simple spatial model can be obtained by integrating a von Mises distribution with parameters $\bm{\theta}=(\mu, \kappa)$ and a geostatistical model, specified by a $K \times K$ correlation matrix $\bm{\Omega}(\rho)$ with entries $\omega_{hk}= \exp \left(-\rho d(h,k)\right), \, \rho > 0$, where $d$ is the 
Euclidean distance between location $h$ and $k$. Under this setting, \eqref{eq:general.model} reduces to the finite-dimensional marginal distribution of a continuous-domain spatial process, indexed by 3 parameters.  The second row of Table \ref{tab:estimates} displays the estimates, where the significant correlation parameter indicates that nonignorable spatial autocorrelation is present, motivating the proposed spatial model.

Geostatistical models are often estimated to perform spatial prediction at unobserved points $s$ of the spatial domain, also known as spatial kriging \citep{banerjee2003hierarchical}. Therefore, we use the proposed model to perform circular kriging.  
Specifically, for each point $s$ of the spatial domain, we first obtain the linear predictor $\hat{z}_s=\bm{\omega}_{s}(\hat{\rho})^{T}\bm{\Omega}(\hat{\rho})^{-1}\bm{z}(\bm{y}; \hat{\mu}, \hat{\kappa})$, where $\bm{\omega}_s(\hat{\rho})^{T}$ includes the estimated correlations
$\hat{\omega}_{sk}=\exp(-\hat{\rho}d(s,k))$ between $s$ and the observed locations. Prediction $\hat{z}_s$ is then transformed into  $\hat{y}_s=z^{-1}(\hat{z}_s)$ to obtain the predicted circular median of the conditional distribution of $y_s$ given the observed data. Figure \ref{fig:kriging} displays such predictions, shaded by grey levels that are proportional to the length of the associated 95\% confidence arc.

\section{Conclusion and discussion}
The paper introduces a flexible class of distributions for multivariate circular data by exploiting Gaussian copulas. 
Differently from previous attempts, the proposed approach does not impose restrictions on the marginal data distribution nor requires overwhelming estimation routines. Inference is based on likelihood maximization, where the starting points of the optimization routine are given following the IFM approach by \cite{kim2016multivariate}.

In addition, conditional distributions can be derived to build regression models as it is illustrated in the two applications. The first one concerns data about animal orientation, for which we assumed identically distributed marginal distributions and a structured correlation matrix. Results show that the latter is able to catch the dynamic of subsequent releases, though unstructured correlation matrices can be embedded in the same framework whenever traditional structures fail to adequately model the data \cite{lopuhaa2023s}. The second application involves the modelling of spatially correlated data about sea current directions recorded in the northern part of the Adriatic Sea. Here, the proposal is used to estimate the spatial correlation and to provide predictions of sea current direction at unobserved locations, with related uncertainty. As in the former application, the correlation structure is fully governed by one parameter, but its estimation still requires the inversion of a $K\times K$ matrix, which may become cumbersome for $K \geq 10^3$. If this is the case, we argue that sparsity-inducing approaches such as \cite{datta2016hierarchical} can be embedded in the proposed framework to solve this issue.

Future development will address the definition of a goodness-of-fit measure following the proposals in \cite{berghaus2017goodness, durocher2017goodness}.


\end{document}